%% file: main.tex
\documentclass{article}
\usepackage{spconf,amsmath,graphicx}
\usepackage{adjustbox}
\usepackage{comment}
\usepackage{caption}
\usepackage{subcaption}
\usepackage{array}
\usepackage[colorlinks=true, pagebackref]{hyperref}
\usepackage{booktabs} 
\usepackage{tabularx}
\usepackage{enumitem}
\usepackage{cite}
\usepackage{mathrsfs}
\usepackage{dsfont}
\usepackage{multirow}

\usepackage{enumitem}
\setlist{nosep, leftmargin=14pt}

\let\OLDthebibliography\thebibliography
\renewcommand\thebibliography[1]{
  \OLDthebibliography{#1}
  \setlength{\parskip}{0pt}
  \setlength{\itemsep}{0pt plus 0.3ex}
}
\title{Multi-Modal Learning Using Physicians Diagnostics for Optical Coherence Tomography Classification}
\begin{document}

\onecolumn 

\begin{description}[labelindent=-1cm,leftmargin=1cm,style=multiline]

\item[\textbf{Citation}]{Y. Logan, K. Kokilepersaud, G. Kwon and G. AlRegib, C. Wykoff, H. Yu, "Multi-Modal Learning Using Physicians Diagnostics for Optical Coherence Tomography Classification," IEEE International Symposium on Biomedical Imaging (ISBI), accepted on Jan. 7 2022.} \\


\item[\textbf{Review}]{Date of accept: 7 Jan 2022} \\


\item[\textbf{Bib}] {@ARTICLE\{Logan2022\_ISBI,\\ 
author=\{Y. Logan, K. Kokilepersaud, G. Kwon and G. AlRegib, C. Wykoff and H. Yu\},\\ 
journal=\{IEEE International Symposium on Biomedical Imaging\},\\ 
title=\{Multi-Modal Learning Using Physicians Diagnostics for Optical Coherence Tomography Classification\}, \\ 
year=\{2022\}\\ 
} \\


\item[\textbf{Copyright}]{\textcopyright 2022 IEEE. Personal use of this material is permitted. Permission from IEEE must be obtained for all other uses, in any current or future media, including reprinting/republishing this material for advertising or promotional purposes,
creating new collective works, for resale or redistribution to servers or lists, or reuse of any copyrighted component
of this work in other works. }
\\
\item[\textbf{Contact}]{\href{mailto:ylogan3@gatech.edu}{ylogan3@gatech.edu}  OR \href{mailto:alregib@gatech.edu}{alregib@gatech.edu}\\ \url{http://ghassanalregib.com/} \\ }
\end{description}

\thispagestyle{empty}
\newpage
\clearpage
\setcounter{page}{1}

\twocolumn

\name{%
\begin{tabular}{@{}c@{}}
Yash-yee Logan$^{\star}$ \qquad 
Kiran Kokilepersaud$^{\ddagger \star}$ \qquad 
Gukyeong Kwon$^{\ddagger \star}$  \qquad 
Ghassan AlRegib$^{\star}$ \qquad \\
{Charles Wykoff}$^{\dagger}$ \qquad 
{Hannah Yu}$^{\dagger}$ 
\thanks{$^{\ddagger}$ Equal contribution.}
\end{tabular}}

\address{$^{\star}$ School of Electrical and Computer Engineering, Georgia Institute of Technology, Atlanta, GA, USA\\
    $^{\dagger}$ Retina Consultants Texas, Retina Consultants of America, Houston, Texas, USA}

\maketitle
\begin{abstract}
\input{abstract}
\end{abstract}
\begin{keywords}
Multi-modal Learning, Diagnostic Attributes, Latent Representation, Autoencoder, OCT
\end{keywords}
\section{Introduction}
\label{sec:intro}

\input{introduction}

\section{Related Works}
\label{sec:format}

\input{relatedWork}

\section{Multi-modal Representations}
\label{sec:multiRep}

\input{multimodalRep}

\section{Attribute Dataset Collection}
\label{sec:typestyle}

\input{attributes}

\section{Experiments}
\label{sec:majhead}

\input{experiments}

\section{Results and discussion}
\label{sec:subhead}

\input{results}

\section{Conclusion}
\label{sec:subsubhead}

\input{conclusion}

\section{Acknowledgments}
\label{sec:acknowledgments}
\vspace{-0.2cm}
This material is based upon work supported by the National Science Foundation Graduate Research Fellowship under Grant No. DGE-1650044.

\section{Compliance with Ethical Standards}
This is a numerical simulation study for which no ethical approval was required.
\vspace{-0.2cm}

\bibliographystyle{IEEEbib}
\bibliography{strings,refs}

\end{document}

%% file: abstract.tex
In this paper, we propose a framework that incorporates experts diagnostics and insights into the analysis of Optical Coherence Tomography (OCT) using multi-modal learning. To demonstrate the effectiveness of this approach, we create a medical diagnostic attribute dataset to improve disease classification using OCT. Although there have been successful attempts to deploy machine learning for disease classification in OCT, such methodologies lack the experts insights. We argue that injecting ophthalmological assessments as another supervision in a learning framework is of great importance for the machine learning process to perform accurate and interpretable classification. We demonstrate the proposed framework through comprehensive experiments that compare the effectiveness of combining diagnostic attribute features with latent visual representations and show that they surpass the state-of-the-art approach. Finally, we analyze the proposed dual-stream architecture and provide an insight that determine the components that contribute most to classification performance.

%% file: introduction.tex
\begin{figure}[t]
\small
\centering
\includegraphics[width=0.83\columnwidth]{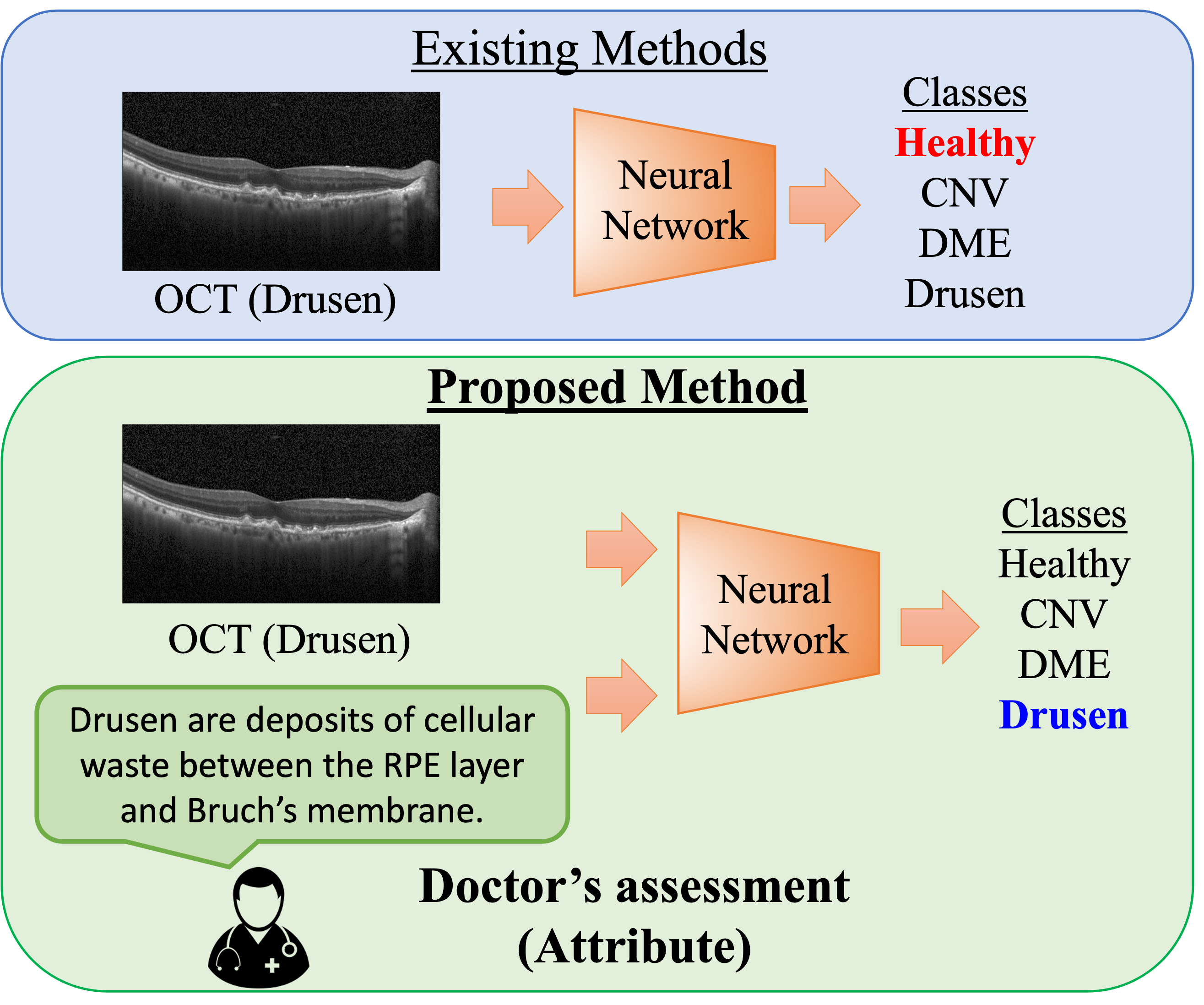}
\caption{High-level overview of proposed multi-modal learning framework using Physicians diagnostics as an attribute. \vspace{-0.6cm}}
\label{fig: high_level}
\end{figure}

Optical coherence tomography (OCT) is a non-invasive imaging technique that provides high-resolution cross-sectional images of the retina. Eye care professionals use it to evaluate the health of the posterior segment of the eye and to detect signs of various vitreoretinal pathologies. The physician's assessment when interpreting an ophthalmic image, provides valuable information that can be a source of knowledge to a model while training. However, current methods are not using this information. Existing approaches for OCT classification typically entail transfer learning with pretrained models \cite{hirano2021universal, yoo2021feasibility}, as shown in Fig.~\ref{fig: high_level}. Though transfer learning often yields better performance than medical experts, very rarely does this transition from research to real clinical settings. This is because transfer learning models make decisions based on unrelated, non-medical, internal weights that are not directly optimised for OCT. This lack of medical intuition behind the decision-making process inspires a lack of trust from the medical community.

In this paper, we aim to mitigate these concerns by considering a physician's assessment as another modality to incorporate with OCT for richer feature representations and improve retinal diagnosis performance. A modality, in this context, is an information channel such as OCT. A multi-modal machine learning problem uses information sourced from multiple communication channels to train models \cite{baltruvsaitis2018multimodal}. Existing multi-modal research with OCT combine it with other medical imaging technologies \cite{leitgeb2018multimodal}. However, combining OCT with diagnostic information is largely unexplored. Diagnostic attributes are text-based and they characterize diseased OCT on patterns a physician would recognize. Combining OCT with diagnostic attributes adds medical context to a model from a multi-modal machine learning perspective. To the best of our knowledge, this is the first attempt to combine ophthalmic imagery with diagnostic attributes for OCT classification.

We highlight the difference between existing and proposed methods for OCT classification in Fig.~\ref{fig: high_level}. Our approach involves fusion of heterogeneous modalities trained from the outset with random weights. The challenge is to transform the data into a meaningful format for the model to join these modalities. Combining the uni-modal information into the same representation space with a dual-stream autoencoder accomplishes this. The dissimilar natures of data make it challenging to optimally exploit any complementarity and redundancies present. Additionally, there is no closed-form relationship between each data type so the ideal mapping of the modalities may not exist \cite{baltruvsaitis2018multimodal}. However, multi-modal learning with diagnostic attributes adds robustness to the feature space as they are not susceptible to variations in image quality like OCT. Also, in practice, networks trained from random initialization improve generalization as they avert rote memorization and emphasize learning over time.  

The diagnostic attributes dataset was acquired by mining a plethora of medical documentation describing attributes indicative of retinal pathologies. During data collection, each attribute noted was chronicled in multiple medical publications. These attributes allow integration of medical insight in the framework to improve OCT classification.
\begin{figure}[t]
\small
\centering
\includegraphics[width=\columnwidth]{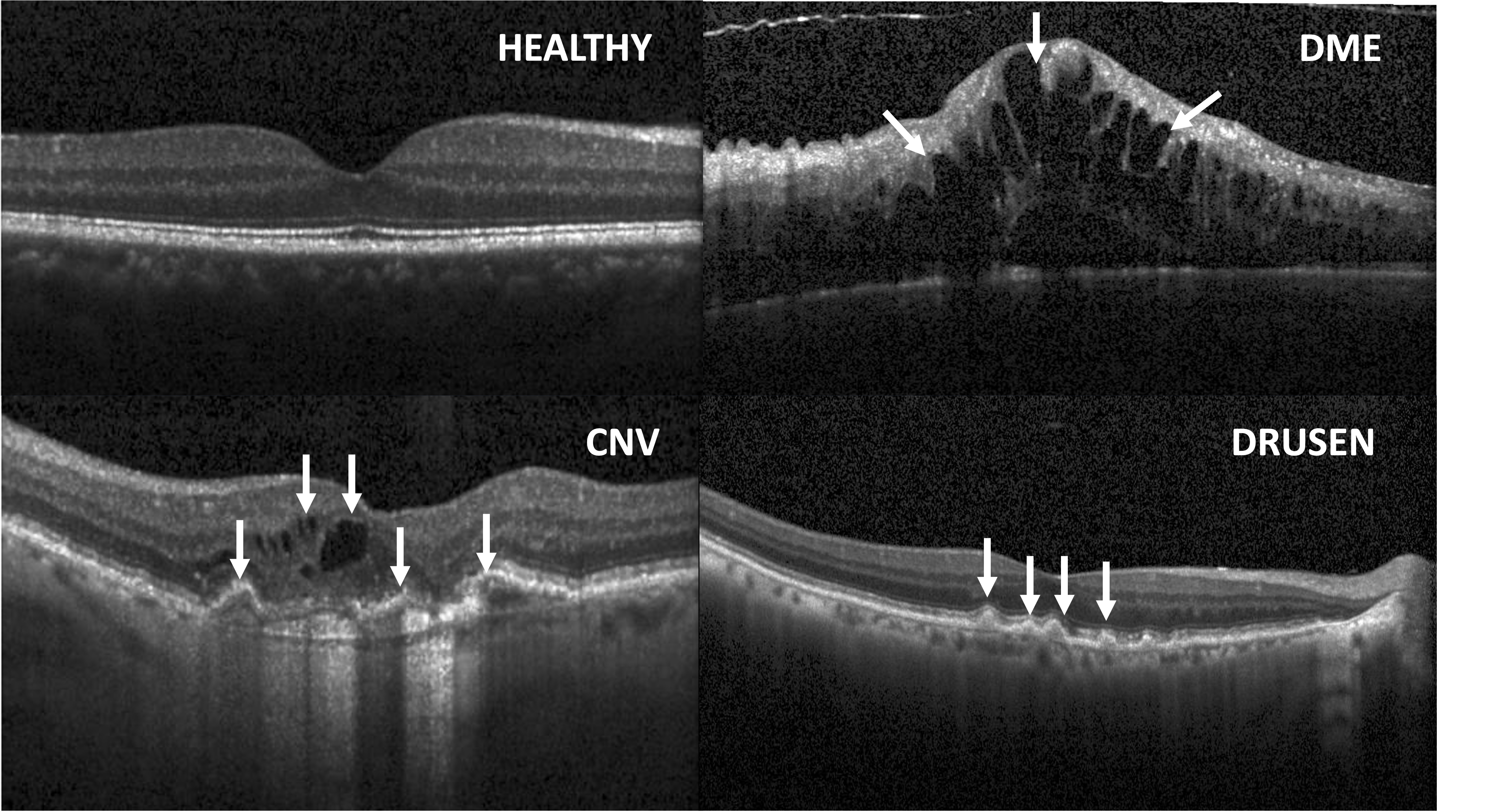}
\caption{A Healthy cross-sectional scan compared to scans with Diabetic Macular Edema (DME), Choroidal Neovascularization (CNV) and Drusen retinal pathologies with arrows highlighting some attributes of each disease.\vspace{-0.8cm}}
\label{fig: examples}
\end{figure}

In this paper, we create a diagnostic attribute dataset consisting of features highly representative of three retinal pathologies. We also demonstrate how these attributes are created. We then propose to characterize these diseases among healthy ophthalmic images through use of this dataset along with OCT. We create a dual-stream architecture and compare it to other architectures trained without diagnostic attributes to show the generalizability of diagnostic attributes for pathology classification. More specifically, we fuse diagnostic attributes with spectral-domain OCT images in order to improve disease classification performance. The contributions of this paper are three fold:
\begin{enumerate}[label=\roman*, leftmargin=0.4cm]
    \item We create an ophthalmic imagery diagnostic attribute dataset with features indicative of three retinal pathologies.
    \item We propose a dual-stream autoencoder to learn joint representations for diagnostic attributes and OCT. 
    \item We show that combining diagnostic attribute features with latent visual representations surpasses the state-of-the-art approach.
\end{enumerate} \vspace{-0.1cm}

%% file: relatedWork.tex

\begin{figure}[t]
\small
\centering
\includegraphics[width=0.95\columnwidth]{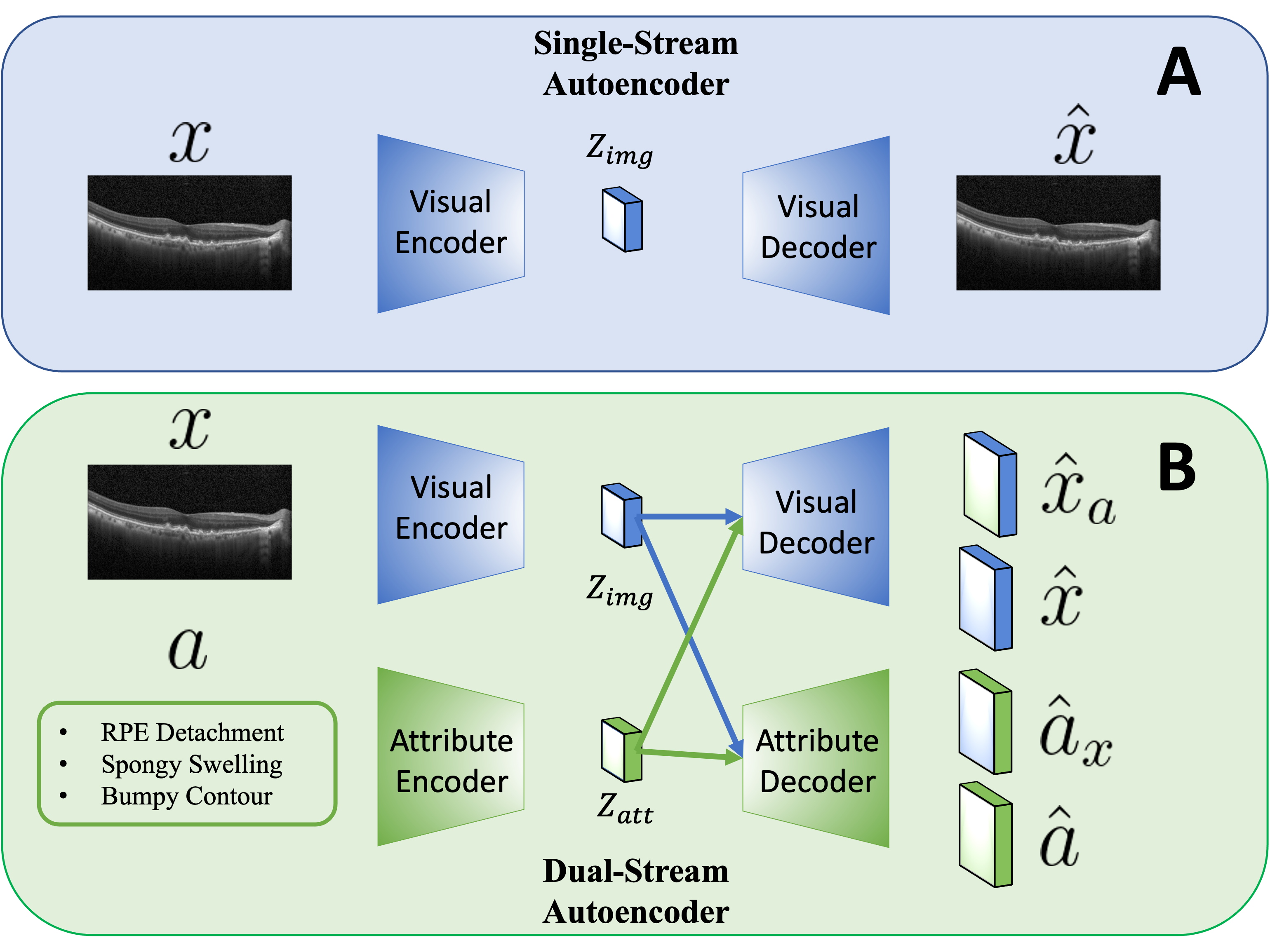}
\caption{Architectures for OCT classification: (A) Single-stream autoencoder (B) Dual-stream autoencoder\vspace{-0.5cm}}
\label{fig: architecture}
\end{figure}
Using multiple modalities for different vision tasks within the medical domain is not new. In the ophthalmological space, this traditionally entails combining modalities from different imaging techniques. For example, the authors of \cite{miri2015multimodal} utilized information jointly between fundus images and OCT scans in order to perform segmentation of the optic disc. More recently, work in \cite{luo2020encoding} takes a retinal image along with a structure map and show that two modalities are better than one. Outside OCT, multiple modalities have also been used to improve classification tasks where representations of MR and PET scans were fused for Alzheimer's diagnosis \cite{liu2014multimodal}. Neither of these utilize a doctor's assessment as an additional modality to integrate more supervision in the learning framework.    

Diagnostic text-based attributes allow an ophthalmologist's assessment to be injected into the model's learning. However, this research task is largely unexplored. A seldom example in \cite{qiao2019mnn} uses clinical textual notes as a method to improve diagnosis performance. However, notes contain irrelevant and possibly inaccurate information that require pre-processing.  Text-based attributes from medical reports are most commonly used with images for multi-modal document retrieval tasks \cite{li2020comparison,hsu2018unsupervised}. The work in  \cite{huang2020fusion} presents an argument for merging medical modalities due to the incomplete picture that a single modality can represent about a patient's condition. However, the methods they present fail to capture direct diagnostic attributes which may offer a complete representation. 

Stacked autoencoders are unsupervised neural models that have also been used to create feature representations for each modality and share them to learn joint a representation~\cite{ngiam2011multimodal}. We use a dual-stream autoencoder to learn joint representations for diagnostic attributes and OCT. In the remaining sections of this paper, we explore how diagnostic attributes impact the the learning framework for retinal disease classification and how our autoencoder architecture improves classification. \vspace{-0.2cm}

%% file: multimodalRep.tex

Autoencoders are the core elements of our proposed multi-modal OCT classification. More specifically, we employ a dual-stream autoencoder, inspired by \cite{ngiam2011multimodal}, that utilizes an encoder and decoder pair for both the visual and attribute input vectors as shown in Fig.~\ref{fig: architecture}. When OCT, $x$, is given to the visual encoder, $f_{v}$, it generates compressed latent features $z_{img} = f_{v}(x)$ that the visual decoder, $g_{v}$, uses to reconstruct its output $\hat{x} = g_{v}(z_{img})$. When diagnostic attributes, $a$, are fed to the attribute encoder, $f_{a}$, it too creates a compact latent variable $z_{att} = f_{a}(a)$ that the attribute decoder, $g_{a}$, uses to reconstruct its output $\hat{a} = g_{a}(z_{att})$. 

The visual $\{{x_1},...,{x_M}\}$ and attribute $\{{a_1},...,{a_M}\}$ inputs have both been projected onto the same latent space and both latent representations are fed back into their own decoder in order to reconstruct their original representation. Minimizing the reconstruction error $\mathcal{L}_{recon}$ allows the reconstructed image and attributes to be less differing from their input. It is defined as: \vspace{-0.2cm}
\begin{align}
\mathcal{L}_{recon} = ||x - \hat{x}||_2 + ||a - \hat{a}||_2
\label{eq:cross}
\end{align}

Second, these representations are then fed into the decoder of the other latent vector. In other words, the visual latent representation undergoes reconstruction in the attribute decoder and vice versa. The cross-reconstruction error, $\mathcal{L}_{cross}$, enforces the model to learn the association between OCT and diagnostic attributes by cross-reconstructing each modality accurately. It is calculated as follows:
\begin{align}
\mathcal{L}_{cross} = ||x - \hat{x}_a||_2 + ||a - \hat{a}_x||_2
\label{eq:cross}
\end{align}

We also impose the latent loss, $\mathcal{L}_{dist}$, which is the cross entropy loss of $l_{2}$ distance between latent representations of OCT and its attribute. This loss enforces alignment between the OCT latent representation and its attribute. $\mathcal{L}_{dist}$ is calculated as follows with $z_{att}^k$ being the latent variable for the $k^{th}$ class attribute:
\vspace{-0.2cm}
\begin{align}
\mathcal{L}_{dist} = -\log\left(\frac{\exp(-||z_{img} - z_{att}||_2)}{\sum_{k=1}^{4}\exp(-||z_{img} - z_{att}^k||_2)}\right)
\label{eq:cross}
\end{align}

Training of the autoencoders is optimised by minimizing the following loss function that has scale factors $\lambda_{cross} = 1$ and $\lambda_{dist}= 0.005$: 
\begin{align}
\mathcal{L} = \mathcal{L}_{recon} + \lambda_{cross}\mathcal{L}_{cross} - \lambda_{dist}\mathcal{L}_{dist} 
\label{eq:cross}
\end{align}

Disease characterization is performed with a classifier that uses latent visual features $z_{img}$ to classify a total of four classes: healthy and three OCT patterns indicative of specific disease states. Its loss is weighted categorical cross entropy, $\mathcal{L}_{class}$. It is calculated as follows with $\hat{y}[class]$ being the classifier's predicted output for the target $class$, $\hat{y}[k]$ being its predicted output for the $k^{th}$ class and $w[class]$ is the inverse frequency weight for the target $class$:
\vspace{-0.4cm}
\begin{align}
\mathcal{L}_{class} = w[class] \left[ -\log \left( \frac{\exp([\hat{y}[class] )}{\sum_{k=1}^{4}\exp(\hat{y}[k])} \right)\right] 
\label{eq:cross}
\end{align}
\vspace{-0.25cm}
\begin{align}
w[class] = \frac{Number \ of \ images \ in \ largest \ class}{Number \ of \ images \ in \ class}
\label{eq:cross}
\end{align}

%% file: attributes.tex
The diagnostic attributes dataset is intended to improve the task of OCT classification. Therefore, all attributes  considered were documented in multiple medical publications~\cite{tuugcu2016imaging,sikorski2013diagnostic, grover2010comparison, montero2005macular} as highly expressive of the three pathologies. For example, by OCT, drusen are visualized as uniform, typically dome-shaped hyperreflectvie collections beneath the retinal pigment epithelium (RPE). We identified 26 visual indicators used by physicians when making diagnoses based on review of OCT images. A subset of these are shown in Table~\ref{tab: attributes}. 

The dataset consists of diagnostic attributes for four classes: healthy, drusen, choroidal neovascularization (CNV) and DME. Each class is associated with an attribute vector with a length equivalent to the number of attributes. For each position within the vector, a 1 represents the presence of that attribute and a 0 represented its absence. The attribute for healthy OCT had 0 for all disease attributes and 1 indicating healthy. These vectors were then associated with each image based on their class and used within the architecture described in Section \ref{sec:multiRep}.
\vspace{-0.1cm}
\begin{table}[h]
\small
\begin{tabular}{p{2cm}|m{6cm}}
    \toprule
 
    \centering \textbf{Pathology} & 
    \centering \textbf{\centering Manifestation in OCT}  \tabularnewline

    \hline
    \midrule 
    \centering DME &
    
    \begin{minipage}[t]{\linewidth}
    \begin{itemize}[nosep,after=\strut, leftmargin=0.4cm]
        \item Reduced intraretinal reflectivity
        \item Hard exudates
    \end{itemize}
    \end{minipage} \\
    
    \hline
    \midrule 
    \centering CNV &
    
    \begin{minipage}[t]{\linewidth}
    \begin{itemize}[nosep,after=\strut, leftmargin=0.4cm]
        \item RPE detachment
        \item Vitreous hemmorhage
    \end{itemize}
    \end{minipage} \\
    \hline
    \midrule 
    \centering Drusen &
    
    \begin{minipage}[t]{\linewidth}
    \begin{itemize}[nosep,after=\strut, leftmargin=0.4cm]
        \item Deposits underneath retina
        \item Bumpy surface
    \end{itemize}
    \end{minipage} \\
    \bottomrule
    \hline
\end{tabular}
\caption{A subset of the diagnostic attributes from each retinal disease.\vspace{-0.3cm}}
\label{tab: attributes}
\end{table}

\begin{table}[t] 
  \begin{center}
  \small
    \begin{tabular}{*{13}{l|c|c|c}}
      \toprule 
      \textbf{Metric} & \textbf{Classifier} & \textbf{Single-Stream} & \textbf{Dual-Stream}  \\
  \hline
      \midrule 
      Average Precision & 0.513 & 0.664  & \textbf{0.703}\\
      Average Recall & 0.499 & 0.672 &  \textbf{0.690}\\
      Average F1 Score & 0.445 & 0.643 & \textbf{0.669}\\
      \bottomrule 
      \hline
    \end{tabular}
    \caption{Metrics comparing model performances. \label{tab:metrics}\vspace{-0.5cm}}
  \end{center}
\end{table}

\begin{table}[h] 
  \begin{center}
  \small
    \begin{tabular}{*{13}{l|c}}
      \toprule 
      \textbf{Metric} & \textbf{$\Delta$ Performance}  \\
  \hline
      \midrule 
      Avg Precision & \textbf{+2.10\%}\\
      Avg Recall &  \textbf{+0.85\%}\\
      Avg F1 Score & \textbf{+1.75\%}\\
      \bottomrule 
      \hline
    \end{tabular}
    \caption{Impact of diagnostic attributes combined with existing pretrained Inception V3 state-of-the-art approach in \cite{kermany2018large}. \label{tab:SOTA} \vspace{-1.2cm}}
  \end{center}
\end{table}
\vspace{-0.8cm}

\begin{table}[h]
\small
\begin{tabular}{ c | c } 
\toprule
\textbf{Ablation} & \textbf{Recall}\\
\hline
\midrule
No regularization & 0.665\\ 
Dropout & 0.675 \\ 
Batch normalization & 0.641\\ 
Batch normalization and dropout & \textbf{0.690}  \\ 
\bottomrule
\hline
\end{tabular}%
\centering
\caption{Ablation study on regularization terms for the dual-stream model \label{fig: abalation}\vspace{-0.5cm}}
\end{table}

\begin{figure}[t]
\small
\label{confMats}
        \begin{subfigure}[b]{0.43\linewidth} \includegraphics[width=\columnwidth]{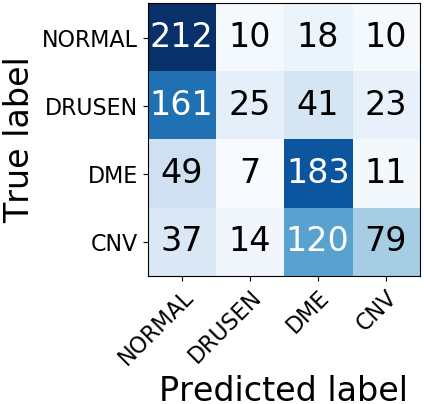}
                \caption*{\vspace{-0.8cm}}
                \label{fig:classifier}
        \end{subfigure}
        \begin{subfigure}[b]{0.5\linewidth} \includegraphics[width=\linewidth]{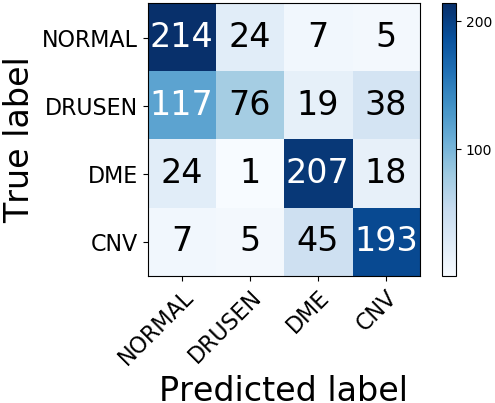}
                \caption*{\vspace{-0.8cm}}
                \label{fig:dual}
        \end{subfigure}%
        \caption{Confusion matrices showing performance of the baseline (left) and dual-stream (right) classification models. \label{confMats}\vspace{-0.7cm}}
\end{figure}
\vspace{-0.2cm}

%% file: experiments.tex
To evaluate the effectiveness of diagnostic attributes for disease classification, we compare the proposed dual-stream autoencoder with two different models. First, we train a classifier directly on OCT as a baseline. The classifier contains two linear layers with relu activation, batch normalization and dropout between. For the second comparison model, we train the single-stream autoencoder as shown in Fig.~\ref{fig: architecture}A and later train the same classifier with its $z_{img}$ features. Comparison between the single-stream and the dual-stream autoencoders will highlight the contribution of diagnostic attributes when the latent representations are utilized for disease classification. All encoders and decoders used are symmetric with two linear layers having a relu activation between. To investigate whether diagnostic attribute features combined with visual features can surpass existing methods, first, we re-implement the Inception V3 architecture, $\phi^{SOTA}$, pretrained on ImageNet that achieved SOTA OCT classification \cite{kermany2018large} in Pytorch. Then, we train a multilayer perceptron ,$\phi^{MLP}$, on attributes and minimize the following objective function inspired by \cite{yan2021positive} to combine diagnostic attributes:  
\vspace{-0.2cm}
\begin{align}
\mathcal{L}_{combo} = \mathcal{L}_{class}^{SOTA} + \mathcal{L}_{class}^{MLP} + \mathcal{L}_{PC}
\label{eq:pct}
\end{align}
\vspace{-0.7cm}
\begin{align}
    \mathcal{L}_{PC} = 1[\hat{y}^{MLP} = y] \frac{1}{2}||\phi^{SOTA}(x_i) - \phi^{MLP}(x_i) ||_2^2 
\end{align}

Finally, an ablation study was conducted to show which regularization terms contribute to learning better joint representations with diagnostic attributes. Fig~\ref{fig: examples} shows examples for each class in the dataset \cite{kermany2018large}. Arrows indicate sample locations for pathology in each disease. The training set consisted of 50280 healthy, 10488 DME, 36345 CNV and 7756 Drusen samples. Validation and test sets consisted of 861 and 250 samples from each class, respectively. During classification, the loss function is weighted by $w[class]$ to counteract the imbalanced dataset. All OCT were resized to $512 \times 128$, normalized and flattened before being given as input. \vspace{-0.2cm}

%% file: results.tex
We report average precision, recall and f1 scores.  The positive impact of combining diagnostic attributes with the SOTA method \cite{kermany2018large} is illustrated in Table \ref{tab:SOTA}.  The results of all other models are in Table~\ref{tab:metrics}. The classifier trained on visual features from the dual-stream autoencoder outperforms both baseline and single-stream models in all metrics. This means there is value in incorporating meaningful knowledge from a doctor's assessment into multi-modal learning framework. Secondly, the baseline experiment having the poorest performance highlights that latent representations are better for the task of OCT disease classification.

Relative to the baseline, the dual-stream model was the best at distinguishing healthy, Drusen, DME and CNV in OCT as shown in Fig.~\ref{confMats}. Drusen is a challenging class to discriminate as it contains the fewest samples in the dataset. Furthermore, the "bumpy" structural nature of Drusen in OCT is sometimes subtle and can be misclassified. Nevertheless, the dual-stream model was still the best at recognizing it.

To investigate which regularization terms of the dual-stream model contributed most to its success, an ablation study was conducted by keeping one regularization component and removing all others. Regularization involved standardizing the mini-batch inputs with batch normalization and randomly discarding some network units via dropout \cite{ioffe2015batch, srivastava2014dropout}. Table~\ref{fig: abalation} shows both regularizations together had the highest impact on model recall. Following this, dropout was the most influential regularization.
\vspace{-0.4cm}

%% file: conclusion.tex
\vspace{-0.3cm}
In this paper, we propose a framework that integrates supervision from medical insights through multi-modal learning. We created a diagnostic attribute dataset to intermix the diagnostic knowledge from ophthalmologists into the machine learning framework. Our experiments validate the effectiveness of diagnostic attributes for OCT classification. In addition, we perform an ablation study to appreciate regularization terms that support learning descriptive joint representations for OCT and diagnostic attributes. Finally, we demonstrate that attributes combined with existing methods surpass SOTA. Encouraged by these results, we will further investigate learning methods to fully utilize diagnostic attributes for the characterization of pathologies.